\documentclass[prl,twocolumn]{revtex4-1}
\usepackage{amsmath}
\usepackage{amssymb}
\usepackage{graphics}
\usepackage{epsfig}
\usepackage{color}
\usepackage{graphicx}
\usepackage{color}
\usepackage{url}

\begin{document}
\title{Communication dynamics in finite capacity social networks} 
\author{Jan O. Haerter$^1$, Bj\o rn Jamtveit$^2$, Joachim Mathiesen$^{1,2}$}
\affiliation{$^1$Niels Bohr Institute, University of Copenhagen, Blegdamsvej 17, DK-2100 Copenhagen, Denmark.$^2$ Physics of Geological Processes, University of Oslo, Norway}


\begin{abstract}
In communication networks structure and dynamics are tightly coupled.
The structure controls the flow of information and is itself shaped by the dynamical process of information exchanged between nodes. In order to reconcile structure and dynamics, a generic model, based on the local interaction between nodes, is considered for the communication in large social networks. In agreement with data from a large human organization, we show that the flow is non-Markovian and controlled by the temporal limitations of individuals. We confirm the versatility of our model by predicting simultaneously the degree-dependent node activity, the balance between information input and output of nodes and the degree distribution. Finally, we quantify the limitations to network analysis when it is based on data sampled over a finite period of time.
\end{abstract}
\maketitle



Limitations on the processing capacities of nodes and links have a profound impact on the flow of information in online communication networks \cite{LK08, Candia08}, the spreading of diseases in human encounter networks \cite{Rocha09}, and in social networks \cite{Borgatti2010,Kitsak2010,C10, DarwinEinstein2005}, where links between interacting individuals
can be highly volatile \cite{Bara2005}. It is often assumed that communication takes place in an unrestrained way on a set of established connections, thereby neglecting, that structure and dynamics are interdependent. Here we consider the evolution of a network where links 
form as a result of non-Markovian interaction between nodes. 
In a time-limited environment, communication demands prioritization which is evident from the analysis of correspondence patterns \cite{DarwinEinstein2005,Malm2009}. Hence, information flow on a network is a result of individuals' choices which are influenced by the state of surrounding nodes. 
In natural \cite{Dunbar1992} and online \cite{Cor2012, Goncalves2011, Ugan2011, Wil2009, Vis2009} social networks, the nodes' activity is a non-trivial function of their degree. The activity level can be quantified by the number of social relationships simultaneously maintained by an individual. This number has been suggested to reflect basic cognitive capabilities of primates \cite{Dunbar1992} and humans \cite{Wil2009, Vis2009, Goncalves2011}. 
Here we model a network of individuals acting under time constraints and compare with a complete dataset of email communication in a large organization. The model is discussed in the context of other communication networks. We predict the information processing capacity of individuals as well as the structure of the network that they form.

\begin{figure}
\centering
\includegraphics[width=.48\textwidth,clip]{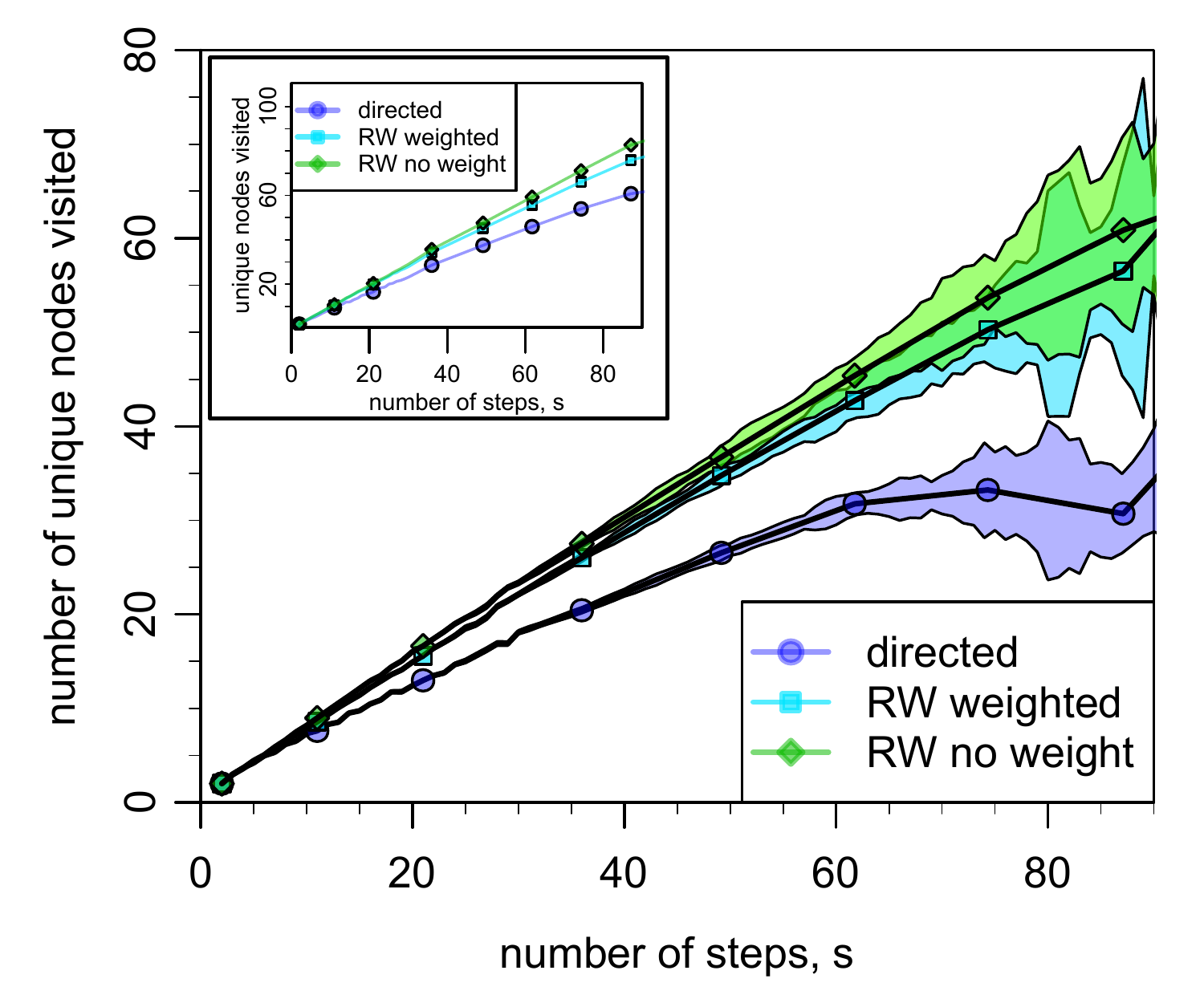}
\caption{Weighted random, unweighted random and directed information flow. The error bars are estimated by bootstrapping. Inset: Similar plot using model data. The quantitative discrepancy between model and data results from the relative dominance of degree-one nodes in the empirical data.}\label{fig:markov}
\end{figure}

We use representative communication data from a large social organization, the University of Oslo. The data comprise a complete time-ordered list of  $2.3\times 10^7$ emails between 5600 employees, 30 000 students and approximately $10^6$ people outside the organization over a period of three months (Sep-Nov 2010). The email content was not recorded and identities of individuals were encrypted.  We limit the influence of unsolicited bulk emails by disregarding those simultaneously sent to more than five recipients. However, the results are not sensitive to the filtering of bulk emails \cite{SM}. Previous work on email data has considered static network structures \cite{Ebel02, Guimera03, Newman02, Eckmann04, Mathiesen10,JJM09}. 

{\it Results --} We show that the communication is non-Markovian by comparing random and directed information flow:
(i) Random flow is given by random walks on the network. The walker follows an empirical time-independent jump-probability $p_{ij}=N_{ij}/\sum_k{N_{ik}}$ from node $i$ to node $j$. The sum is taken over all nodes and $N_{ij}$ is the number of emails sent from $i$ to $j$ during the timespan of the data. 
(ii) Directed flow is given by the chronological email exchange. Starting from a random node $i$, we wait for $i$ to send an email, say to $j$. We then jump to $j$ and wait for the next message $j$ sends either back to $i$ or to a new node $k$. Repeating this, we obtain a finite trajectory within the timespan of the data. The number of unique nodes visited by the directed and random flow as function of the number of jumps are compared by averaging over trajectories originating from all nodes (Fig.~\ref{fig:markov}). On average, directed flow visits relatively fewer nodes than random flow, indicating a significant correlation between sent and received messages. 

\begin{figure}[htb]
\includegraphics[width=.48\textwidth]{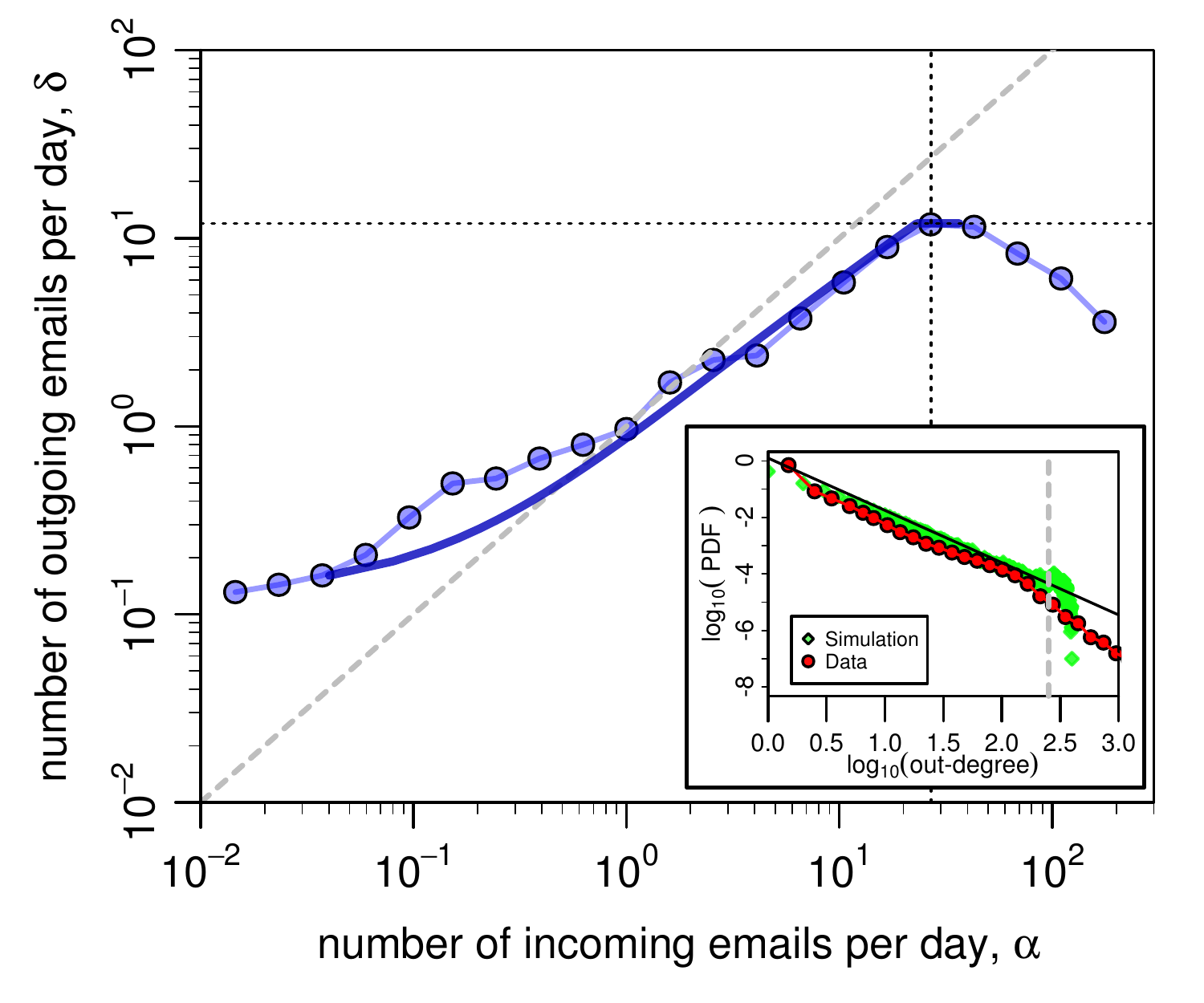}\vspace{-.0cm}\\
\caption{Average number of messages sent per message received. Observational data is marked by "$\circ$". The solid line is a best fit by Eq.~(\ref{eq:delta}). The dotted lines mark the peak and the dashed diagonal line shows $\delta=\alpha$. Inset: out-degree distribution for model and empirical data. The dashed line denotes the scale-break $k_{sb}\simeq 250$. Mean degree is $5.4$ (Twitter data yields a mean degree of $8.8$ and a similar exponent for the degree distribution~\cite{dechoud}). Note the double-log scales.}
\label{fig:sent_vs_rec}
\end{figure}

Our model requires nodes to perform a trade-off between replying to others and initiating new conversations. Specifically, consider $\mathcal{N}$ nodes, each initially connected to one other node.
The nodes have a limited capacity and can send a maximum of $N_{max}$ messages in a timestep $\Delta t=1$ day. The dynamics follows from three possible actions for a node $i$ of out-degree $k_i$: 

(a) $i$ processes received emails and if $i$ has sent less than $N_{max}$ messages, any received email is replied to with a probability proportional to the sender's degree. Emails not replied to within $\Delta t$ are subsequently deleted. In total, $\delta_2$ replies are sent by this action.

(b) If less than $N_{max}$ emails have been sent in (a), the remaining capacity $N_{max}-\delta_2$ is available for sending messages, called $\delta_1$, to previously established contacts. The probability of sending a message to a contact is given by a constant $r_{ini}$. Hence, granted sufficient capacity on average $r_{ini}\cdot k_i$ messages are initiated by $i$. Nodes with low $k_i$ will generally not reach their full capacity. 

(c) Nodes establish new contacts by sending requests with a probability $r_{req}$. The probability that a request is sent to a node $j$ is proportional to the degree of $j$, $k_j$. A link is established between $i$ and $j$, if $j$ in the next timestep according to (a) replies to $i$. In reality, contacts might as well be established by face-to-face encounters, i.e. via channels not recorded explicitly in our data.

The total number of messages $\delta$ sent by a node in $\Delta t$ is the sum $\delta\equiv \delta_0+\delta_1+\delta_2$. Analogously, messages received by a node in the same timestep are termed $\alpha\equiv\alpha_0+\alpha_1+\alpha_2$. Nodes have an average lifetime $\tau$ and are therefore removed from the network with a probability $\Delta t/\tau$. For every node removed, a new node with a single random connection to an existing node is introduced. 
$\tau$ is estimated to be $5.8$ years from the known mean email user turnover time in the organization.  
The parameters $r_{ini}$, $r_{req}$ and $N_{max}$ are determined below.

\begin{figure}[h]
\includegraphics[width=.48\textwidth]{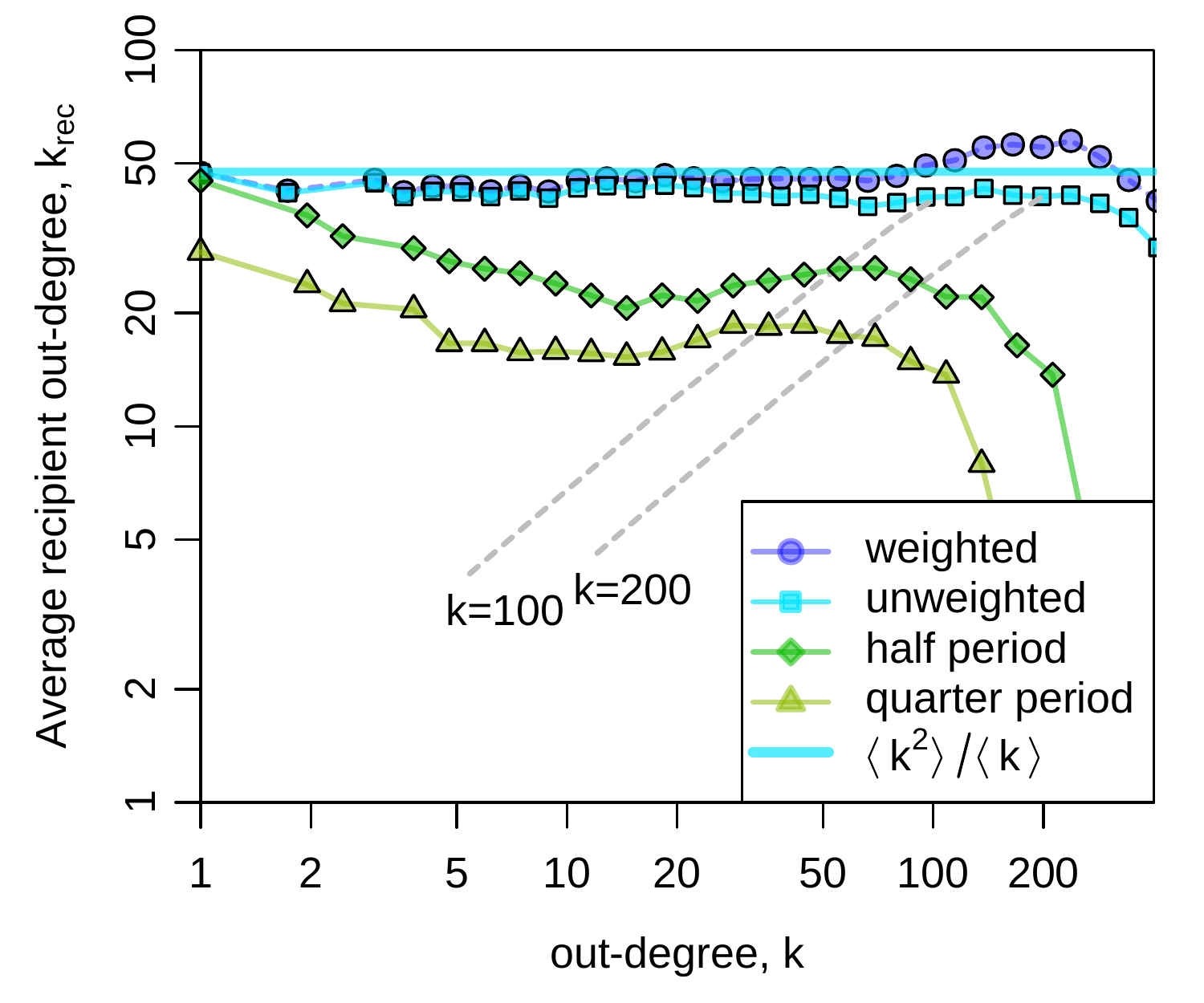}\vspace{-.0cm}
\caption{Mean recipient degree as function of degree ($\Box$) and weighted by the number of messages sent to recipients ($\circ$). The horizontal line shows $\langle k^2\rangle/\langle k\rangle$.  The curves marked by "$\diamond$" and "$\triangle$" are analogous to the unweighted case but for half, respectively, one quarter of the observational period. Dashed lines show projection of nodes with two values of $k$ for a varying observation window.
Note the double-log scale.}
\label{fig:rec_degree_vs_k}
\end{figure}

According to (c), a link is established between $i$ and $j$ if one of the nodes sends a message to the other and receives a reply. The probability, $P_{ij}$, that a message is sent from $i$ to $j$ in $\Delta t$ is proportional to $k_j$, 
\begin{equation}
P_{ij}=\frac{r_{req}k_j}{\sum_{\ell\neq i}k_\ell}\approx\frac{r_{req}k_j}{\mathcal{N}\langle k\rangle}\;,\label{eq1}
\end{equation}
where we in the approximation assume that $k_i\ll\sum_\ell k_\ell$.
According to (a), the mean number of requests that $j$ receives during a timestep is proportional to $r_{req}$ and $k_j$. The probability for $j$ to reply to a request from nodes of degree $k$ is proportional to $\beta k n(k)$, where $\beta$ is a constant and $n(k)$ is the number of nodes with degree $k$. The number of replies written by $j$ is the product of Eq.~(\ref{eq1}) and the integral over nodes
\begin{equation}
\frac{r_{req} k_j}{\mathcal{N} \langle k\rangle}\int \beta~k~n(k)~\mathrm{d}k=\beta r_{req} k_j.
\label{eq:prob_send_reply}
\end{equation}
Since nodes reply to requests and therefore establish new links with a probability proportional to the sender degree, $k n(k)$, the mean degree $k_c$ of a node's contacts is $k_c\equiv\int k^2 n(k) \mathrm{d}k / \int k n(k)\mathrm{d}k=\langle k^2\rangle/\langle k\rangle$, a number generally larger than the mean degree $\langle k\rangle$ (Fig.~\ref{fig:rec_degree_vs_k}).

Consequently the average degree-increase of nodes of degree $k$ per timestep becomes $r(k)\Delta t\equiv 2\beta r_{req}k\Delta t$. The factor of 2 reflects the symmetry of sending and replying. The rate of losing links is inversely proportional to $\tau$, $d\equiv k/\tau$. Hence, the net degree-growth rate becomes
$\Delta k/\Delta t= k\cdot r_0$, 
where $r_0\equiv\left(2\beta r_{req}-\tau^{-1}\right)$. As long as a node has sufficient capacity to reply to all requests its degree increases approximately exponentially, $k(t)\sim \exp(r_0 t)$. 


The degree distribution follows from the consideration that during $\Delta t$, a fraction of nodes $n(k)$ of degree $k$ changes their degree,  $r_0\left[(k-1)n(k-1)-kn(k)\right]$, and a fraction $1/\tau$ is removed. A continuum-limit approximation yields 
\begin{equation}
 \frac{\partial n(k)}{\partial t}=-r_0\left[k\frac{\partial n(k)}{\partial k}+n(k)\right]-\frac{n(k)}{\tau}\;.
\label{eq:time_evol}
\end{equation}
The steady-state solution has the form $n(k)=n(1)\cdot k^{-\gamma}$, where $\gamma\equiv (1-1/2\beta r_{req}\tau)^{-1}$. The constant $n(1)$ is fixed by integrating Eq.~(\ref{eq:time_evol}) over $k$ and by demanding that the total number of nodes $\mathcal{N}=\int dk~n(k)$ be constant. This yields $n(1)=\mathcal{N}(\gamma-1)$. The condition $0<n(1)<\mathcal{N}$ bounds the power-law exponent: $1<\gamma<2$. The data yield $\gamma\simeq 1.85$ (Fig.~\ref{fig:sent_vs_rec} inset).


So far we have assumed that nodes have infinite capacity. As a node's degree increases, it receives more messages and this assumption becomes invalid. 
Consider the number of messages received by $i$ per timestep. Contact requests from other nodes amount to $\alpha_{0}\equiv r_{req}k_i/\langle k \rangle$ messages. The senders of these messages are drawn from a distribution $n(k)/\mathcal{N}$. The probability for $i$ to receive a message from its contacts is proportional to $r_{ini}$ and $k_i$, hence $\alpha_1\equiv r_{ini}\cdot k_i$. Analogously, as defined in (a), $i$ issues $\delta_0\equiv r_{req}$ requests to recipients distributed according to $\rho_1(k)$ (where $\rho_\ell(k)\equiv k^\ell n(k)/\int {k'}^\ell n(k')dk'$) due to the weighting of probabilities by the recipient degree. In the same timestep $i$ sends $\delta_1=\alpha_1$ messages to its contacts. Finally we consider back-and-forth communication. For every message sent by $i$ to $j$, a response is returned with a probability $\beta k_i$ (Eq.~\ref{eq:prob_send_reply}). In steady-state, the number of messages sent is identical for all timesteps and therefore $i$ receives 
\begin{equation}
\alpha_2\equiv \beta k_i \left(\delta_0+\delta_1+\delta_2\right)
\label{eq:alpha2} 
\end{equation}
replies to messages sent in the previous timestep. $\delta_2$ is the number of messages $i$ sends in response to messages received from others which again is a sum over contributions from the actions (a)-(c):
\begin{equation}
 \delta_2\equiv \beta\left(\alpha_0\langle k\rangle_{\rho_0}+\alpha_1\langle k\rangle_{\rho_1}+\alpha_2\langle k\rangle_{\rho_{\alpha_2}}\right)\;.\label{eq5}
\end{equation}
The terms on the right are respectively, requests from any node in the network (distributed as $\rho_0$), messages from existing contacts (distributed as $\rho_1$), and back-and-forth messages (distributed as $\rho_{\alpha_2}$). Each iteration of back-and-forth communication acts as a shift in the distribution of recipients relative to the distribution of senders $\mathcal{F}\rho_l\equiv\beta\rho_{l+1}$. The distribution $\rho_{\alpha_2}$ accounts for all high-order shifts.
To close the equations for $\alpha_2$ and $\delta_2$, we use that the reply probability for each iteration is reduced by a factor $\beta$ to approximate $\rho_{\alpha_2}\simeq\rho_2$. Inserting Eq.~(\ref{eq:alpha2}), $\alpha_0$ and $\alpha_1$ in Eq.~(\ref{eq5}) yields 
$\delta_2=\beta\left(\alpha_0\langle k\rangle +\alpha_1 k_c+\beta k_i k_c(\delta_0+\delta_1)\right)/f(k_i)$
where we introduce $f(k_i)\equiv 1-\beta^2k_i k_c\leq 1$. Summing over $\delta_{0}$, $\delta_{1}$ and $\delta_{2}$ we get 
\begin{equation}
 \delta=r_{req}+r_{ini}k_i+\frac{\beta k_i}{f(k_i)}\left(r_{req}+r_{ini}k_c+\beta k_c(r_{req}+r_{ini}k_i)\right)\;.
\label{eq:delta}
\end{equation}
Here the first three terms (referred to as $\delta_{<}$) are messages sent to recipients selected according to $\rho_1$ and with mean degree $k_c$. The other terms, $\delta_{>}$, are messages to recipients distributed according to the higher order distribution $\rho_2$ which has a mean $k_c^*\equiv\langle k^3\rangle/\langle k^2\rangle>k_c$ and contribute significantly only for large $k_i$. 
The mean of the weighted recipient degree (weighted by number of messages received) is $k_{rec}^w\equiv k_c\delta_{<}/\delta+k_c^*\delta_{>}/\delta $, which departs from $k_c$ when $\delta_{>}$ becomes appreciable (Fig.~\ref{fig:rec_degree_vs_k}).
For low $k_i$ ($k_i=1$), the ratio of sent to received messages becomes $\delta/\alpha\simeq (r_{req}+r_{ini})/(r_{req}/\langle k\rangle+r_{ini})>1$. 
Conversely, $\delta/\alpha=1$ when $k_i=\langle k\rangle$, hence an average node has a ``balanced'' email account. When $k_i$ becomes larger than $\langle k\rangle$, $i$ will increasingly receive requests and responses to its messages (Fig.~\ref{fig:sent_vs_rec}).

\begin{figure}
\includegraphics[width=.48\textwidth]{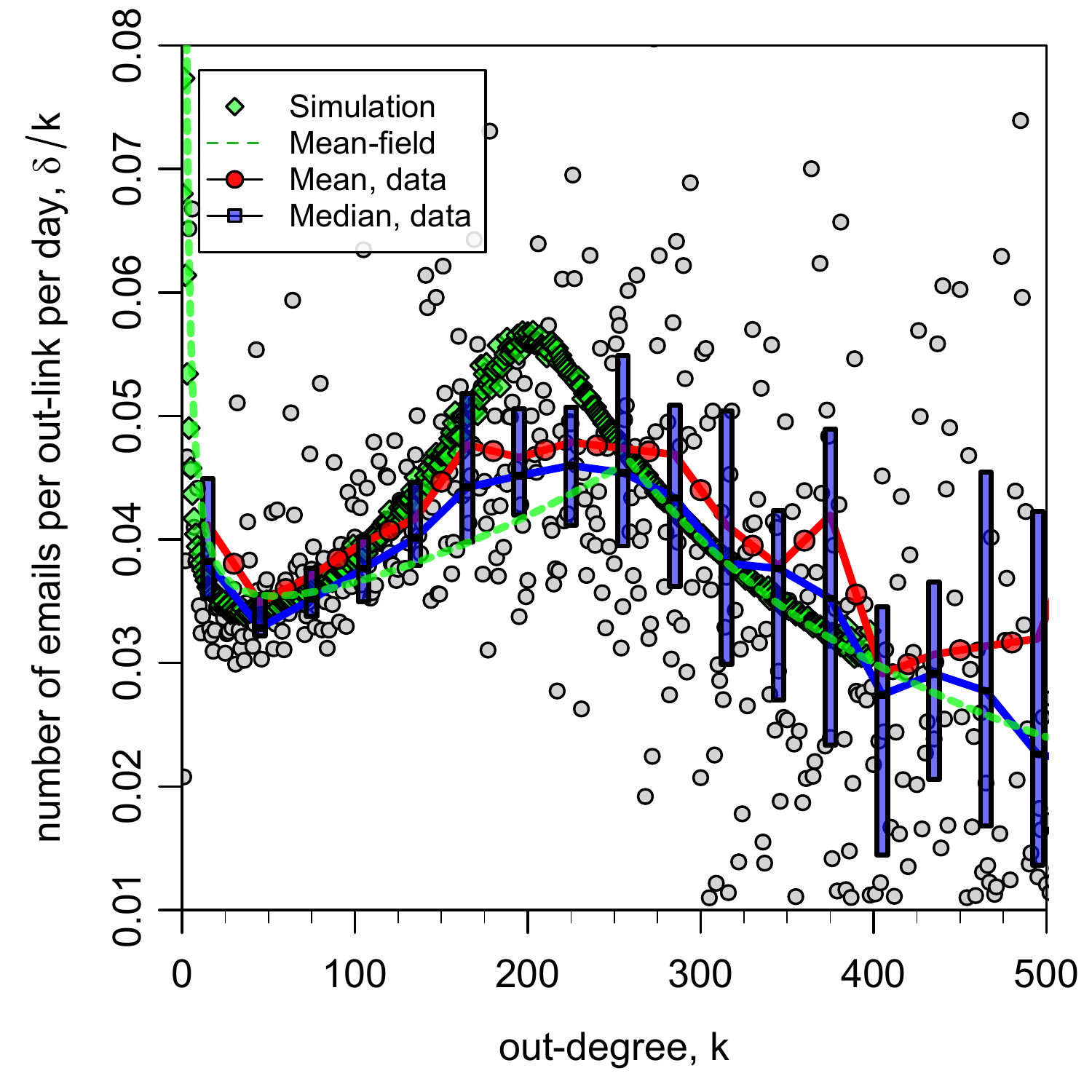}
\caption{Average number of emails sent per link per day. Gray circles represent the average activity of all users of a certain out-degree and the red (blue) lines represent coarse grained mean (median) values in the real communication network; boxes mark upper and lower quartiles. Best fit with the model (simulation) is shown by the green lines (diamonds). At small $k$, $f(k)\simeq 1$ (Eq.~\ref{eq:delta}) and $\delta/k$ is a superposition of a term $\sim k$ due to the final quadratic term and a decaying term $\sim k^{-1}$ from the constant. 
At $k>k_D$, nodes limited to $N_{max}$ messages per day, hence $\delta/k\sim N_{max}/k$. 
}\label{fig:dunbar}
\end{figure}

The {\it Dunbar number} $k_{D}$ is the degree where $\delta$ reaches the capacity limit ($\delta=N_{max}$) and $\delta/k$ is maximal. The scale break in the degree distribution ($k_{sb}\simeq 250$), Fig.~\ref{fig:sent_vs_rec} (inset), and $k_D\simeq 230$, Fig.~\ref{fig:dunbar}, nearly coincide. In fact $k_{sb}$ is related to $k_D$ because nodes beyond $k_D$ have a reduced probability to form new links. To determine $k_{sb}$, consider the evolution of the nodes' degree in the limit where all capacity is used for replying, hence $\delta_1=0$. 
Using that $\delta_0\ll \delta_2$, we get $\delta\approx\delta_2= N_{max}$ which in turn yields  $k_{sb}=\beta^{-1}N_{max}f(k_{sb})\left(r_{req}+r_{ini}k_c\right)^{-1}$. $k_{sb}$ is found by solving this implicit equation. $k_D$ then follows from Eq.~(\ref{eq:delta}).

The parameters $r_{ini}=0.023$, $r_{req}=0.13$ and $N_{max}=12$ are determined by the data in Fig.~\ref{fig:sent_vs_rec}. From $r_{req}$ and $\gamma$ we obtain $\beta\approx 0.004$. Larger $N_{max}$ increases the limit of $\delta$. $r_{req}$ is constrained by the offset at low $\alpha$ and $r_{ini}$ effects the skewness of the curve which follows from analysis of Eqs.~(\ref{eq:alpha2}) and (\ref{eq:delta}).
Fig.~\ref{fig:dunbar} shows the model prediction of $\delta/k_i$ and the corresponding email data. 
We complement our analysis with numerical computations. Using a large number of nodes, $\mathcal{N}=10,000$, we iterate actions (a)-(c) until steady-state is reached. While the mean-field prediction (Figs.~\ref{fig:rec_degree_vs_k} and \ref{fig:dunbar}) is close to the numerical solution, some differences exist, e.g.~at small $k$, $\rho(k)$ is not a strict power-law in the numerical solution due to the discreteness of $k$. 
Further, the simulation gives a smooth peak in $\delta/k$ (Fig.~\ref{fig:dunbar}) which is narrower than in the empirical data. This is due to slight overestimation of the repeated back-and-forth communication between well-connected nodes ($k\approx 200$) relative to the data.  
We have also simulated the information flow (Fig.~\ref{fig:markov}) and achieve similar results. Finally, the average local clustering coefficient of the empirical and simulated networks is relatively small, $\approx 0.04$ for both (similar clustering coefficient $\approx 0.06$ \cite{dechoud} and $k_D\approx 150$ to $200$ have been reported for other communication networks \cite{Vis2009, Goncalves2011, chun}). We further checked the robustness of the model to variations \cite{SM}.

%

{\it Discussion --} The data were recorded over three months and the communication network is therefore a finite-time projection of the real network.
The projection reduces the number of links. More active links will more likely persist through the projection than less active links. Fig.~\ref{fig:rec_degree_vs_k} shows the mean recipient degree $k_{rec}$ as function of the sender degree $k_i$ for three observation time intervals. Consider again Eq.~(\ref{eq:delta}) and remember that recipients of the $\delta_{<}$ ($\delta_{>}$) messages are distributed as $\rho_1$ ($\rho_2$). When observing only a single day, the probability for an out-link between $i$ to $j$ not to be active is $P_{ij}(\Delta t)\equiv 1-\delta_{<}k_j/k_ck_i-\delta_{>}k_j/k_c^*k_i$. For $d$ days we obtain ${P}_{ij}(d\Delta t)= {P}_{ij}(\Delta t)^d$. To produce the projected curves in Fig.~\ref{fig:rec_degree_vs_k}, ${P}_{ij}(d\Delta t)$ is applied to both axes, $k$ and $k_{rec}(k)$. 
Averaging w.r.t.~all recipients $j$ (distributed as $\rho_1$), the projected sender out-degree becomes $k_i^{(d)}\equiv k_i\langle 1- {P}_{ij}^d\rangle_{\rho_1}$. Similarly one can consider the projection of the mean recipient degree leading to a similar reduction in the degree for finite-time data.
For example, consider the data for the quarter period ($d\approx 23$) in Fig.~\ref{fig:rec_degree_vs_k}. We have ${P}_{ij}(\Delta t)^d\simeq (1-r_{ini})^d$ and therefore $k_i^{(d)}/k_i<1/2$ hence less than half the links persist. 


{\it Concluding remarks --} The finite capacity of agents in social networks induces an upper limit on the number of possible interactions~\cite{Wil2009,Vis2009,Goncalves2011,Cor2012}.
We propose a comprehensive model that reconciles structure and dynamics of networks with finite capacity agents that dynamically form or lose links. 
In agreement with a complete set of email data and results from other social networks \cite{Cor2012,dechoud}, our model predicts a scale-free degree distribution up to a distinct scale-break induced by the capacity limit. 
Further, as agents gain importance in the network, the per-link-activity first increases with node-degree, peaks at intermediate degrees and declines at large degrees. 
The model and data therefore support the hypothesis of a general limit on the number (150-250) of active social relations that an individual can maintain \cite{Dunbar1992} and is in agreement with empirical observations on social networks \citep{Goncalves2011, chun}.

\section*{Acknowledgments}
This study was supported by the Danish National Research Foundation through the Center for Models of Life and by Physics of Geological Processes, a Center of Excellence at the University of Oslo. 
Email data from the University of Oslo were collected with the help and support of Ingar Vindenes and Knut Borge at the Univ. of Oslo Center for Information Technology (USIT).
 
%

\end{document}